\documentclass{article}
\usepackage{spconf,amsmath,graphicx}
\usepackage{amsfonts}
\usepackage{amsmath,bm}
\usepackage{adjustbox}
\usepackage{algorithmicx}
\usepackage{algpseudocode}
\usepackage[ruled]{algorithm}
\usepackage{graphicx}
\usepackage{caption}
\usepackage{graphicx,times,amsmath} 
\usepackage{caption}
\usepackage{subcaption}
\usepackage[rightcaption]{sidecap}
\usepackage{caption}
\usepackage{multirow}
\usepackage{hhline}
\usepackage{amsmath}
\usepackage{epstopdf}
\usepackage{tabularx}
\usepackage{tikz}
\usetikzlibrary{matrix}
\usepackage{mathtools, nccmath}
\usepackage{pst-node}

\usepackage{makecell}

\usepackage{lipsum}

\newcommand\blfootnote[1]{%
  \begingroup
  \renewcommand\thefootnote{}\footnote{#1}%
  \addtocounter{footnote}{-1}%
  \endgroup
}

\graphicspath{ {images/} }

\DeclarePairedDelimiter\floor{\lfloor}{\rfloor}
\usepackage{alphalph}

\title{L\MakeLowercase{ite}HAR: Lightweight Human Activity Recognition from WiFi Signals \\ with Random Convolution Kernels}

\name{Hojjat Salehinejad, Member, IEEE, and Shahrokh Valaee, Fellow, IEEE}
\address{Department of Electrical \& Computer Engineering, University of Toronto, Toronto, Canada \\
\textit{hojjat.salehinejad@mail.utoronto.ca, valaee@ece.utoronto.ca}}

\usepackage{fancyhdr}
\begin{document}
\newcommand*{\img}{%
  \includegraphics[
    width=\linewidth,
    height=20pt,
    keepaspectratio=false,
  ]{example-image-a}%
}

\maketitle

\begin{abstract}
Anatomical movements of the human body can change the channel state information (CSI) of wireless signals in an indoor environment. These changes in the CSI signals can be used for human activity recognition (HAR), which is a predominant and unique approach due to preserving privacy and flexibility of capturing motions in non-line-of-sight environments. Existing models for HAR generally have a high computational complexity, contain very large number of trainable parameters, and require extensive computational resources. This issue is particularly important for implementation of these solutions on devices with limited resources, such as edge devices. In this paper, we propose a lightweight human activity recognition (LiteHAR) approach which, unlike the state-of-the-art deep learning models, does not require extensive training of a large number of parameters. This approach uses  randomly initialized convolution kernels for feature extraction from CSI signals without training the kernels. The extracted features are then classified using Ridge regression classifier, which has a linear computational complexity and is very fast. LiteHAR is evaluated on a public benchmark dataset and the results show its high classification performance with a much lower computational complexity in comparison with the complex deep learning models.   
\end{abstract}
\begin{keywords}
Channel state information, random convolution kernels, human activity recognition, time series.
\end{keywords}

\section{Introduction}
\label{sec:intro}
 \blfootnote{Accepted for presentation at IEEE ICASSP 2022. \\ \copyright 2022 IEEE. Personal use of this material is permitted. Permission
 from IEEE must be obtained for all other uses, in any current or future
media, including reprinting/republishing this material for advertising or
 promotional purposes, creating new collective works, for resale or
 redistribution to servers or lists, or reuse of any copyrighted
 component of this work in other works.}
 
 The WiFi technology, based on the IEEE 802.11n/ac standards~\cite{yousefi2017survey}, uses 
 Orthogonal Frequency Division Multiplexing (OFDM) which decomposes the spectrum into multiple subcarriers with a symbol transmitted over each subcarrier.
 {\it Channel state information} (CSI) reflects how each subcarrier 
 is affected through the wireless communication channel.
 CSI in a Multiple-Input Multiple-Output (MIMO) setup has spatial diversity due to using multiple antennas as well as frequency diversity due to using multiple subcarriers per antenna. Each CSI sample, measured at the baseband, is a vector of complex variables where the size of the vector is the number of subcarriers in the OFDM signal.
 
 It has been shown that changes in the environment affect the CSI and hence 
 human activities can be recognized by studying the CSI variations~\cite{yousefi2017survey, wang2015understanding}.
 Anatomical movements of the human body can change the reflection of WiFi signals, hence change the CSI in the environment.  
 By a proper study of the CSI variations certain human activities can be detected. 
 Besides the availability of WiFi signals in homes and most indoor environments, the privacy issue makes the use of CSI for HAR an attractive proposition. 
 Unlike cameras, WiFi-based activity recognition preserves privacy of users and does not require a line-of-sight access. 
 
 Some of the early models for human activity recognition (HAR) are based on random forest (RF), and the hidden Markov model (HMM). It has been shown that these methods have lower performance than deep learning approaches such as LSTM~\cite{yousefi2017survey}, SAE~\cite{gao2017csi}, ABLSTM~\cite{chen2018wifi}, and WiARes~\cite{cui2021device}. The WiAReS~\cite{cui2021device} model is an ensemble of multi-layer perceptron (MLP) networks, convolution neural networks (CNNs), RF, and support vector machine (SVM). Its performance is close to that of the ABLSTM~\cite{chen2018wifi} method. Unfortunately, the computational complexity of this approach is not reported and the codes are not available.
 
Generally, deep learning approaches have a large number of trainable parameters, require tremendous training data, need major hyper-parameter tuning, and are resource hungry in training and inference\cite{salehinejad2017recent}. Training deep learning models with limited-imbalanced data is challenging~\cite{wang2016training} and augmentation methods can partially reduce the overfitting of these models~\cite{salehinejad2018synthesizing,salehinejad2018image}. Deployment of these solutions generally requires access to graphical processing units (GPUs) which make their implementation expensive on resource-limited devices without access to cloud resources. However, it is possible to design structural-efficient networks~\cite{luo2017thinet} and prune unnecessary parameters to shrink the models~\cite{han2015deep,salehinejad2021edropout}.

\begin{figure*}[!t]
\centering
\captionsetup{font=footnotesize}
\includegraphics[width=0.79\textwidth]{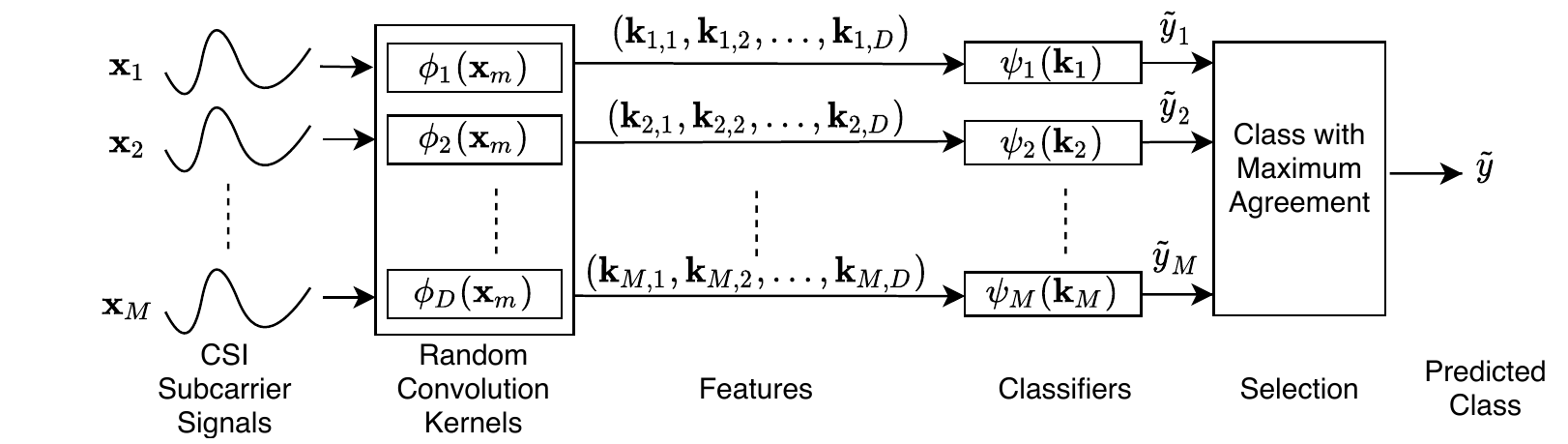}
\caption{Proposed model for activity recognition on the input CSI subcarrier amplitude signals $(\mathbf{x}_{1},...,\mathbf{x}_{M})$.}
\vspace{-4mm}
\label{fig:csirocket_model}
\end{figure*}

It is possible to use a large number of random convolution kernels, without training them, for feature extraction from time series as proposed in Rocket~\cite{dempster2020rocket}. This is a fast and accurate time series classification approach which has showed significant performance improvement in classification tasks for various applications such as driver’s distraction detection using electroencephalogram (EEG) signals~\cite{tan2020detecting} and functional near infrared spectroscopy signals classification~\cite{andreu2021single}. 

This paper proposes a novel approach for HAR based on Rocket~\cite{dempster2020rocket}. Unlike the deep learning approaches~\cite{yousefi2017survey, gao2017csi,chen2018wifi,cui2021device}, our method does not require training of a large number of parameters and is computationally very light, hence called {\it lightweight human activity recognition} (LiteHAR). LiteHAR also does not require a GPU setup and can be implemented on local devices without cloud access\footnote{The codes and more details of the experiment setup are available at: \textit{https://github.com/salehinejad/LiteHAR}}.



\section{L\MakeLowercase{ite}HAR Model}
Different steps of the proposed LiteHAR model are presented in Fig.~\ref{fig:csirocket_model}. The steps are described in detail next. 

\subsection{Input Signals}
\begin{figure}[!t]
\centering
\captionsetup{font=footnotesize}
\includegraphics[width=0.4\textwidth]{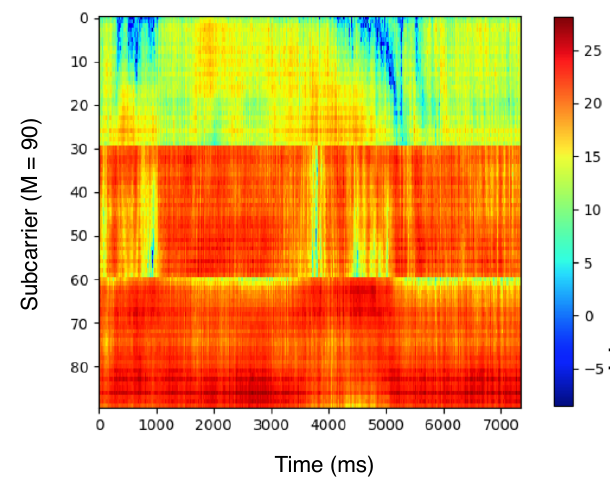}
\caption{A sample spectrogram of CSI signals per subcarrier over time.}
\vspace{-4mm}
\label{fig:spect}
\end{figure}

Let $\{(\mathbf{X}_{1},y_{1}),...,(\mathbf{X}_{N},y_{N})\}$ represent a set of $N$ training samples where $\mathbf{X}_{n}=(\mathbf{x}_{n,1},...,\mathbf{x}_{n,M})$ is the CSI amplitude signal of $M$ subcarriers over time and $y_{m}$ is the corresponding activity label.
As an example, for a MIMO receiver with three antennas and $30$ subcarriers per antenna, $(\mathbf{x}_{1},...,\mathbf{x}_{30})$, $(\mathbf{x}_{31},...,\mathbf{x}_{60})$, and $(\mathbf{x}_{61},...,\mathbf{x}_{90})$ represent the CSI amplitude signals of subcarriers on antenna one to three, respectively, as demonstrated in Fig.~\ref{fig:spect}. 
Unlike most activity recognition methods (e.g.~\cite{yousefi2017survey,chen2018wifi}), we do not perform any major pre-processing on the input signals.

\subsection{Feature Extraction}
Let $(\mathcal{\phi}_{1},...,\mathcal{\phi}_{D})$ represent a set of convolution operations where the kernels $1,...,D$ are randomly initialized based on Rocket~\cite{dempster2020rocket} as follows:
\begin{itemize}
\item Length: Randomly selected from $\{7, 9, 11\}$ with equal probability;
\item Weights: Randomly sampled from a Normal distribution $\mathcal{N}(0,1)$;
\item Bias: Randomly sampled from a uniform distribution $\mathcal{U}(-1,1)$;
\item Dilation: Randomly sampled from an exponential scale $\floor*{2^{a}}$, where $a\sim \mathcal{U}(0, log_{2}\frac{l_{input}-1}{l_{kernel}-1})$,  $l_{kernel}$ is the kernel length, and $l_{input}$ is the length of the input signal;
\item Padding: Applied randomly with equal probability; 
\item Stride: Set to one for all kernels.
\end{itemize}

The convolution outputs are then represented by two values per kernel, the proportion of positive values ($ppv$) and the maximum value ($max$)~\cite{dempster2020rocket}. Hence, the extracted features for each CSI subcarrier signal $\mathbf{x}_{m}$ is $(\mathbf{k}_{m,1},...,\mathbf{k}_{m,D})$, where ${\mathbf{k}_{m,d}=(ppv_{m,d},max_{m,d})}$. As Fig.~\ref{fig:csirocket_model} shows, for $M$ given subcarriers, $M$ feature vectors are generated. An advantage of this feature representation approach is mapping a variable-length time series to a fixed-length feature vector, which eliminates the padding of different-length signals.

\begin{table*}[]
\captionsetup{font=footnotesize}
\caption{Classification performance results and training and inference time of RF~\cite{yousefi2017survey}, HMM~\cite{yousefi2017survey}, LSTM~\cite{yousefi2017survey}, SAE~\cite{gao2017csi}, ABLSTM~\cite{chen2018wifi}, and the proposed LiteHAR method with $10$-fold cross-validation.}
\label{T:accuracy}
\centering
\begin{adjustbox}{width=0.95\textwidth}
\begin{tabular}{c|cccccccc|cccc}
 Method & \makecell{Lie\\ down} & Fall  & Walk & Run & \makecell{Sit \\down} & \makecell{Stand \\up} & \makecell{Pick \\up} & Avg. & \makecell{Total Training \\ Time (Sec.)}&  \makecell{Total Inference \\ Time (Sec.)}& \makecell{Inference Time \\ per Sample (Sec.)} \\ \hline
RF & 0.53     & 0.60  & 0.81  & 0.88     & 0.49  & 0.57 & -  & 0.65&
6.09 & 0.016 & 3.8e-5\\
HMM & 0.52     & 0.72  & 0.92  & 0.96     & 0.76  & 0.52 & - & 0.73& 0.22  & 0.029 & 5.2e-4\\ 
SAE & 0.84     & 0.84  & 0.95  & 0.83     & 0.84  & 0.88 & -  &0.86& 1788.28 & 0.23 & 5.4e-4 \\ \hline
LSTM & 0.95     & 0.94  & 0.93  & 0.97     & 0.81  & 0.83 & -& 0.90& 5168.86 & \textbf{4.39} & \textbf{0.010} & $\downarrow$ $90\%$\\ 
ABLSTM & 0.96 & \textbf{0.99}  & 0.98  & 0.98     &\textbf{ 0.95}  & \textbf{0.98} & - & \textbf{0.97}&
13007.20 &  6.86 & 0.016\\ 
LiteHAR & 0.92     & 0.93  & \textbf{0.99}  & \textbf{0.99}     & 0.86  & 0.94 & -  & 0.93 &
 \textbf{157.8} &  5.46 & 0.013\\ \hline
\makecell{LiteHAR \\(7 classes)}& 0.90     & 0.90  & \textbf{0.99}  & 0.95     & 0.82  & 0.94 &  \textbf{0.93} &  0.91&
 171.5&  5.46 & 0.013\\
\end{tabular}
\end{adjustbox}
\vspace{-4mm}
\end{table*}

\subsection{Classifier}

In the proposed model, we train a classifier $\psi(\cdot)$ per subcarrier. Therefore, for the $M$ extracted feature vectors $(\mathbf{k}_{1},...,\mathbf{k}_{M})$, a classifier bank $\big(\psi_{1}(\mathbf{k}_{1}),...,\psi_{M}(\mathbf{k}_{M})\big)$ is trained. The predicted activity class from the bank of classifiers is $\tilde{\mathbf{y}}=(\tilde{y}_{1},...,\tilde{y}_{M})$. 

The predicted class for an input $\mathbf{X}_{n}$ with target activity class $y_{n}$ over all the subcarriers is extracted by voting in $\tilde{\mathbf{y}}$ as
\begin{equation}
    \tilde{y}=\underset{c}{\text{argmax}}\Big\{\sum_{m=1}^{M}\mathbf{1}{[\tilde{y}_{m},1]},...,\sum_{m=1}^{M}\mathbf{1}{[\tilde{y}_{m},C]}\Big\},
    \label{eq:voting}
\end{equation}
where $c\in(1,...,C)$ and $C$ is the number of activity classes and $\mathbf{1}_{[\tilde{y}_{m},c]}$ is the indicator function such that $\mathbf{1}[\tilde{y}_{m},c]=1$ if $\tilde{y}_{m}=c$ and $\mathbf{1}[\tilde{y}_{m},c]=0$ if $\tilde{y}_{m}\neq c$. From (\ref{eq:voting}), it is obvious that ${\sum_{c=1}^{C}\sum_{m=1}^{M}\mathbf{1}{[\tilde{y}_{m},c]}=M}$. If $\tilde{y}=y_{n}$ the model has made a correct prediction.

Commonly, the ridge regression classifier is used in Rocket, which is a very simple and significantly fast classifier, and uses generalized cross-validation to determine appropriate regularization~\cite{dempster2020rocket}. This classifier is used as $\psi_{m}(\cdot)$ in this paper, but one is not limited to this classifier.

\section{Experiments}
\label{sec:experiments}

\subsection{Data}
The experiments were conducted on the CSI dataset\footnote{\textit{\text{https://github.com/ermongroup/Wifi\_Activity\_Recognition}}} provided in~\cite{yousefi2017survey}. The CSI data was collected at a receiver with three antennas and $30$ sub-carriers at a sampling rate of $1$kHz. The length of each collected sample is $20$ Seconds. This dataset has $7$ activity classes which are \textit{Run}, \textit{Pick up}, \textit{Lie down}, \textit{Fall}, \textit{Sit down}, \textit{Stand up}, and \textit{Walk}, collected in an indoor environment. 
Most proposed methods in the literature (e.g.~\cite{yousefi2017survey,chen2018wifi,cui2021device}) have been evaluated on $6$ activity classes (i.e. the \textit{Pick up} activity class has been excluded) of the dataset. For the sake of comparison, LiteHAR is evaluated on these $6$ classes as well as on the entire dataset (i.e. $7$ activity classes).

\subsection{Setup}
The proposed LiteHAR model is implemented in Python using Numba\footnote{\textit{http://numba.pydata.org/}} high performance compiler and parallel and lightweight pipelining\footnote{\textit{https://joblib.readthedocs.io/en/latest/}}. Our codes are available online\footnote{\textit{https://github.com/salehinejad/LiteHAR}}. The input CSI signals are down-sampled to $500$Hz and normalized by subtracting the mean and dividing by the $l_2$-norm. The number of random kernels is ${D=10,000}$~\cite{dempster2020rocket}. Regularization strength is set to 10 evenly spaced numbers on a log-scale in the range $(-3,3)$.
The average results of $10$ independent runs are reported and the training dataset is shuffled in each run. A computational setup similar to~\cite{chen2018wifi} was used.

\subsection{Classification Performance Analysis}

Table~\ref{T:accuracy} shows the classification accuracy results of the RF~\cite{yousefi2017survey}, HMM~\cite{yousefi2017survey}, LSTM~\cite{yousefi2017survey}, SAE~\cite{gao2017csi}, ABLSTM~\cite{chen2018wifi}, and the proposed LiteHAR model. The confusion matrices of the top three models are presented in Table~\ref{T:six_class}.

For the experiments on $6$ activity classes, ABLSTM has the highest classification performance at $97\%$ with the best overall accuracy for three activity classes. The accuracy of LiteHAR model is slightly lower than ABLSTM, where it has achieved the best overall performance for two activity classes. However, LiteHAR has a training time of $157.8$ sec while the training time for ABLSTM is about $82\times$ more than  LiteHAR. In inference, LiteHAR is $0.003$ sec faster than ABLSTM. Note that current version of LiteHAR is implemented for parallel processing on CPUs. However, a GPU implementation can further accelerate the training and inference of the model.

LiteHAR has achieved a performance of $91\%$ in classification of $7$ activity classes with an accuracy of $93\%$ for the \textit{Pick up} activity class. Adding this class has  dropped the accuracy of LiteHAR by $2\%$ from the $6$-class model. The confusion matrix of LiteHAR for all the activity classes is presented in Table~\ref{T:seven_class}.

\begin{table}[]
\captionsetup{font=footnotesize}
\caption{Confusion matrix of the LSTM~\cite{yousefi2017survey}, ABLSTM~\cite{chen2018wifi}, and proposed LiteHAR methods over the \textit{Lie down}, \textit{Fall}, \textit{Walk}, \textit{Run}, \textit{Sit down}, and \textit{Stand up} activity classes. Rows may not sum to one due to the rounding artifact.}
\label{T:six_class}
\centering
\begin{adjustbox}{width=0.44\textwidth}
\begin{subtable}[t]{0.6\textwidth}
\caption{LSTM}
\begin{tabular}{cc|cccccc}
\multicolumn{8}{c}{Predicted}    \\ 
\multirow{7}{*}{\rotatebox[origin=c]{90}{Actual}} &          & Lie down & Fall & Walk & Run  & Sit down & Stand up \\ \cline{2-8} 
                        & Lie down & \textbf{0.95}     & 0.01 & 0.01 & 0.01 & 0.0      & 0.02     \\ 
                        & Fall     & 0.01     & \textbf{0.94} & 0.05 & 0.0  & 0.0      & 0.0      \\  
                        & Walk     & 0.0      & 0.01 & \textbf{0.93 }& 0.04 & 0.01     & 0.01     \\
                        & Run      & 0.0      & 0.0  & 0.02 & \textbf{0.97} & 0.01     & 0.0      \\ 
                        & Sit down & 0.03     & 0.01 & 0.05 & 0.02 & \textbf{0.81}     & 0.07     \\ 
                        & Stand up & 0.01     & 0.0  & 0.03 & 0.05 & 0.07     & \textbf{0.83}     \\  
\end{tabular}
\end{subtable}
\end{adjustbox}
\begin{adjustbox}{width=0.44\textwidth}
\begin{subtable}[t]{0.6\textwidth}
\caption{ABLSTM}
\begin{tabular}{cc|cccccc}
\multicolumn{8}{c}{Predicted}    \\ 
\multirow{7}{*}{\rotatebox[origin=c]{90}{Actual}} &          & Lie down & Fall & Walk & Run  & Sit down & Stand up \\ \cline{2-8} 
                        & Lie down & \textbf{0.96}     & 0.0  & 0.01 & 0.0  & 0.02     & 0.02     \\  
                        & Fall     & 0.0      & \textbf{0.99} & 0.0  & 0.01 & 0.0      & 0.0      \\ 
                        & Walk     & 0.0      & 0.0  & \textbf{0.98} & 0.02 & 0.0      & 0.0      \\ 
                        & Run      & 0.0      & 0.0  & 0.02 &\textbf{ 0.98} & 0.0      & 0.0      \\  
                        & Sit down & 0.01     & 0.01 & 0.01 & 0.0  & \textbf{0.95}     & 0.02     \\ 
                        & Stand up & 0.01     & 0.0  & 0.0  & 0.0  & 0.01     & \textbf{0.98 }    \\ 
\end{tabular}
\end{subtable}
\end{adjustbox}
\begin{footnotesize}*Rows may not sum to one due to the rounding artifact, from \cite{chen2018wifi}.
\end{footnotesize}

\begin{adjustbox}{width=0.44\textwidth}
\begin{subtable}[t]{0.6\textwidth}
\caption{LiteHAR}
\begin{tabular}{cc|cccccc}
\multicolumn{8}{c}{Predicted}    \\ 
\multirow{7}{*}{\rotatebox[origin=c]{90}{Actual}} &          
                                    & Lie down & Fall & Walk & Run  & Sit down & Stand up \\ \cline{2-8} 
                        & Lie down  & \textbf{0.92}      & 0.02  & 0.0  & 0.0   & 0.02     & 0.01     \\ 
                        & Fall      & 0.01      & \textbf{0.93}  & 0.0  & 0.0  & 0.02      & 0.02      \\  
                        
                        & Walk     & 0.0       & 0.0   & \textbf{0.99 } & 0.01  & 0.0      & 0.0      \\ 
                        
                        & Run      & 0.0       & 0.0   & 0.01  & \textbf{0.99}  & 0.0      & 0.0      \\ 
                        
                        & Sit down & 0.03      & 0.05  & 0.0  & 0.0   & \textbf{0.86}    & 0.06     \\ 
                        
                        & Stand up & 0.01      & 0.03   & 0.0   & 0.0   & 0.01     & \textbf{0.94}     \\ 
\end{tabular}
\end{subtable}

\end{adjustbox}
\end{table}

\begin{table}[]
\captionsetup{font=footnotesize}
\caption{Confusion matrix of the proposed LiteHAR model over the \textit{Lie down}, \textit{Fall}, \textit{Walk}, \textit{Run}, \textit{Sit down}, \textit{Stand up}, and \textit{Pick up} activity classes. Rows may not sum to one due to the rounding artifact.}
\label{T:seven_class}
\centering
\begin{adjustbox}{width=0.495\textwidth}
\begin{tabular}{cc|ccccccc}
\multicolumn{9}{c}{Predicted}    \\ 
\multirow{7}{*}{\rotatebox[origin=c]{90}{Actual}} &          
                                    & Lie down & Fall & Walk   & Run    & Sit down & Stand up & Pick up\\ \cline{2-9} 
                        & Lie down  & \textbf{0.90}     & 0.01  & 0.0  & 0.0   & 0.06     & 0.01    &  0.01 \\  
                        
                        & Fall      & 0.01      & \textbf{0.90}  & 0.0  & 0.0    & 0.0       & 0.07   & 0.02  \\  
                        
                        & Walk     & 0.0       & 0.0   & \textbf{0.99}  & 0.01   & 0.0       & 0.0     & 0.0\\  
                        
                        & Run      & 0.0       & 0.0   & 0.02  & \textbf{0.95}   & 0.0       & 0.0     & 0.0\\
                        
                        & Sit down & 0.06      & 0.05  & 0.0    & 0.0      & \textbf{0.82}   & 0.05   & 0.01 \\ 
                        
                        & Stand up & 0.0      & 0.03   & 0.0   & 0.01   & 0.01     & \textbf{0.94}    & 0.01\\ 
                        & Pick up & 0.01      & 0.02   & 0.0   & 0.0    & 0.01     & 0.01    &  \textbf{0.93}\\
\end{tabular}
\end{adjustbox}
\end{table}

\begin{figure*}[!t]
\centering
\captionsetup{font=footnotesize}
\includegraphics[width=1\textwidth]{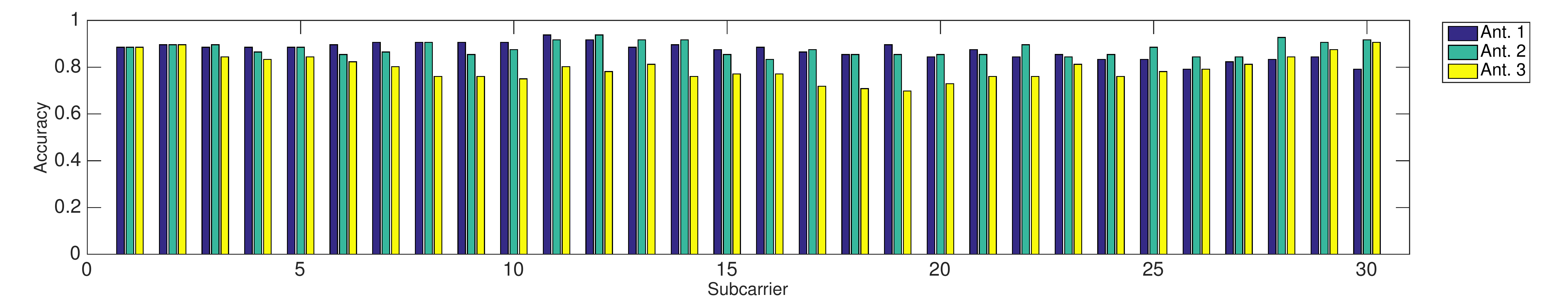}
\caption{Classification accuracy performance of the proposed model per subcarrier and per antenna (Ant.) over all activity classes. Overall accuracy is $92\%$.}
\label{fig:subcarrier_acc}
\end{figure*}

\subsection{Computational Complexity of LiteHAR}
The LiteHAR model has two parts. The first part is applying random convolution transforms for feature extraction, which has a computational complexity of $O_{T}=O(D\cdot N \cdot l_{input})$, where $l_{input}$ is the length of the time series~\cite{dempster2020rocket}.
This complexity is a linear function of the number of kernels. The second part is the Ridge regression classifier, which has a complexity of $O_{R}=O(N^{2}\cdot D)$ when $N<D$,~\cite{dempster2020rocket}.

\subsection{Spatial Diversity Analysis}
The MIMO system with OFDM modulation offers spatial-frequency diversity in CSI data collection. Fig.~\ref{fig:subcarrier_acc} shows the classification performance of $\big(\psi_{1}(\mathbf{k}_{1}),...,\psi_{M}(\mathbf{k}_{M})\big)$ per subcarrier ($30$ subcarrier/antenna) of each antenna ($3$ antennas), for a single run over all activity classes with an overall accuracy of $92\%$.
Fig.~\ref{fig:subcarrier_acc} shows that not all the antennas and subcarriers contribute to the performance of the model and some are redundant or have negative impact. From a spatial perspective, antenna $3$ has a lower overall accuracy than the other antennas (subcarriers $3$ to $30$). However, all subcarriers in antennas $1$ and $2$ have a competitive performance. Hence, one may detect and prune the redundant/destructive antennas/subcarriers from the voting mechanism in $(\ref{eq:voting})$. This approach can enhance the classification performance of a model. During our experiments, we have observed that training LiteHAR with captured signal from the first two antennas can increase its classification performance about $1\%$.

\section{Conclusions}
\label{sec:conclusion}
WiFi-based solutions for human activity recognition (HAR) offer privacy and non-line-of-sight activity detection capabilities. Most recent proposed methods, which have achieved high classification performance on benchmark datasets, use complicated deep learning solutions with very large number of trainable parameters. In order to have an affordable and practical HAR solution, particularly for implementation on resource-limited devices in a local setup, both classification accuracy and computational complexity of the model should be considered. In this paper, we have proposed a lightweight human activity recognition (LiteHAR) solution, which has a very competitive classification performance in comparison with the state-of-the-art methods and has very low computational complexity. Unlike most deep learning solutions, LiteHAR does not require training of a large number of parameters and can be implemented on resource limited devices without GPU access. 


\section{Acknowledgment}
This work was partially supported by the Mobile AI Lab established
between Huawei Technologies Co. LTD Canada and The Governing Council of the University of Toronto.
\bibliographystyle{IEEEbib}
\bibliography{strings,mybibfile}

\end{document}